\documentstyle[preprint,aps,epsf,floats,tighten]{revtex} 

\begin{document}


\preprint{\vbox{ \hbox{UCSD/PTH 98-21} \hbox{hep-ph/9806237} \hbox{} }}
\title{\bf NEW STRATEGIES FOR EXTRACTING ${V_{ub}}^{^{\, \#}}$ }
\footnotetext{$\!^{\#}$Talk presented at the XXXIIIrd Rencontres de Moriond: 
Electroweak Interactions and Unified Theories, Les Arcs, Savoie, France, 
March 14--21 1998.} 

\author{\large Zoltan Ligeti}

\address{\vspace{.2cm} \it Department of Physics,
  University of California San Diego \\ La Jolla, CA 92093--0319, USA}

\maketitle

\vspace{-0.2truecm}
\begin{abstract}%
The prospects for a precise and model independent determination of $|V_{ub}|$
from inclusive and exclusive semileptonic $B$ decays are reviewed.   

\end{abstract}

\baselineskip=18pt

\section{Introduction}

The next generation of $B$ decay experiments will test the flavor sector of the
standard model with high precision.  The basic approach is to determine the
elements of the CKM matrix using different methods and then check for the
consistency of these results.  In practice this amounts to determining the
sides and angles of the unitarity triangle from $CP$ conserving decays and from
$CP$ asymmetries, respectively.  For these checks to be meaningful, a precise
and model independent determination of the $b\to u$ CKM matrix element,
$|V_{ub}|$, is very important.

$CP$ violation has only been observed in kaon decay arising from $K^0-\bar K^0$
mixing.  This observation can be accommodated in the three generation standard
model using the otherwise free parameter $\delta$ in the CKM matrix, but this
description of $CP$ violation has not yet been tested.  Moreover, to explain
the baryon asymmetry of the universe, other sources of $CP$ violation are
needed \cite{MaTr}.  Many extensions of the standard model have new particles
with weak scale masses, and could give large contributions to $CP$ asymmetries
or flavor changing neutral current processes (like $K^0-\bar K^0$ mixing,
$B^0-\bar B^0$ mixing, $B\to K^* \gamma$, etc.).

At the present time, the magnitude of $V_{ub}$ is determined by comparing
experimental results on the endpoint region of the electron spectrum in
inclusive $B$ decays with phenomenological models \cite{incl}, or by comparing
experimental results on $B\to\rho\,\ell\,\bar\nu$ and $B\to\pi\,\ell\,\bar\nu$
with phenomenological models and lattice QCD results \cite{excl}.  These two
approaches yield remarkably consistent determinations of $|V_{ub}|$, but have
large theoretical uncertainties.\footnote{$V_{ub}$ is one of the least
precisely known parameters of the standard model.  The other poorly known CKM
matrix element, $V_{td}$, is related to $V_{ub}$ through the unitarity
triangle.}

To reduce these uncertainties, it will be advantageous to consider different
observables.  The problem is that there is almost no overlap between quantities
sensitive to $|V_{ub}|$ which can be reliably calculated theoretically and
those which can be measured experimentally.  Here I discuss two proposals which
bridge this gap to some extent.  In Section II the possibility of extracting
$|V_{ub}|$ from the hadron invariant mass spectrum in inclusive semileptonic
$B$ decay is reviewed.  Section III concerns extracting $|V_{ub}|$ from a
double ratio of form factors in exclusive semileptonic $B$ and $D$ decays to
$\rho$ and $K^*$.  Section IV contains some conclusions.

\section{$V_{\lowercase{ub}}$ from inclusive $B$ decays}

The traditional method for extracting $|V_{ub}|$ from experimental data
involves a study of the electron energy spectrum in inclusive charmless
semileptonic $B$ decay \cite{incl}.  In the $B$ rest frame, electrons with
energies in the endpoint region $E_e>(m_B^2-m_D^2)/2m_B$ must arise from $b\to
u$ transitions.  There has been considerable theoretical progress recently in
our understanding of inclusive semileptonic $B$ decay \cite{CGG,Bigi,MaWi},
based on the use of the operator product expansion (OPE) and heavy quark
effective theory.  At leading order in the $\Lambda_{\rm QCD}/m_b$ expansion
the $B$ meson decay rate is equal to the $b$ quark decay rate.  There are no
nonperturbative corrections of order $\Lambda_{\rm QCD}/m_b$.  In the electron
endpoint region our calculational ability is lost, since the size of this
region $m_D^2/2m_B\simeq330\,$MeV is comparable to $(m_B-m_b)/2$.  An infinite
set of higher order terms in the OPE, which extend the endpoint from $m_b/2$ to
$m_B/2$, yield singular contributions to ${\rm d}\Gamma/{\rm d}E_e$ that are
equally important integrated over such a small endpoint region.  

In the future, it may be possible to determine $|V_{ub}|$ from a comparison of
the measured hadronic invariant mass spectrum in the region $s_H<m_D^2$ with
theoretical predictions \cite{FLW,BDU,old}.  Here $s_H=(p_B-q)^2$, where $p_B$
is the $B$ meson four-momentum, and $q=p_e+p_{\bar\nu}$ is the sum of the
lepton four-momenta.  An obvious advantage to studying this quantity rather
than the lepton energy spectrum is that most of the $B\to X_u\,e\,\bar\nu$
decays are expected to lie in the region $s_H<m_D^2$, while only a small
fraction of the $B\to X_u\,e\,\bar\nu$ decays have electron energies in the
endpoint region.  Both the invariant mass region, $s_H<m_D^2$, and the electron
endpoint region, $E_e>(m_B^2-m_D^2)/2m_B$, receive contributions from hadronic
final states with invariant masses that range up to $m_D$.  However, for the
electron endpoint region the contribution of the states with masses nearer to
$m_D$ is kinematically suppressed. This region is dominated by the $\pi$ and
the $\rho$ in the ISGW model \cite{ISGW}, with higher mass states making only a
small contribution.  The situation is very different for the low invariant mass
region, $s_H<m_D^2$.  Now all states with invariant masses up to $m_D$
contribute without any preferential weighting towards the lowest mass ones.  In
the ISGW model the $\pi$ and the $\rho$ mesons comprise only about a quarter of
the $B$ decays to states with $s_H<m_D^2$.  Consequently, it is much more
likely that the first few terms in the OPE will provide an accurate description
of $B$ semileptonic decay in the region $s_H<m_D^2$ than in the endpoint region
of the electron energy spectrum.  In fact, from a theoretical point of view,
the cut $s_H<m_D^2$ provides the optimal kinematical separation between
inclusive $b\to u$ and $b\to c$ decays.  A modest cut on the electron energy,
which will probably be required experimentally for the direct measurement of
$s_H$ via the neutrino reconstruction technique, will not destroy this
conclusion.  

To begin with, consider the contribution of dimension three operators in the
OPE to the hadron mass squared spectrum in $B\to X_u\,e\,\bar\nu$ decay. 
This is equivalent to $b$ quark decay and implies a result for ${\rm
d}\Gamma/{\rm d}E_0\,{\rm d}s_0$ (where $E_0=p_b\cdot(p_b-q)/m_b$ and
$s_0=(p_b-q)^2$ are the energy and invariant mass of the strongly interacting
partons arising from the $b$ quark decay) that can easily be calculated using
perturbative QCD up to order $\alpha_s^2\beta_0$.  Even at this leading order
in the OPE there are important nonperturbative effects that come from the
relation between the $b$ quark mass and the $B$ meson mass,
$m_B=m_b+\bar\Lambda+{\cal O}(\Lambda_{\rm QCD}^2/m_b)$.  The most
significant effect comes from $\bar\Lambda$, and it relates the hadronic
invariant mass $s_H$ to $s_0$ and $E_0$ via
\begin{equation}
s_H = s_0 + 2 \bar\Lambda E_0 + \bar\Lambda^2 \,.
\end{equation}
Changing variables from $(s_0, E_0)$ to $(s_H, E_0)$ and integrating $E_0$
over the range
\begin{equation}
\sqrt{s_H} - \bar\Lambda < E_0 <
  {1\over 2m_B}\, (s_H - 2\bar\Lambda m_B + m_B^2),
\end{equation}
gives ${\rm d}\Gamma/{\rm d}s_H$, where $\bar\Lambda^2<s_H<m_B^2$.  Feynman
diagrams with only a $u$-quark in the final state contribute at $s_0=0$, which
corresponds to the region $\bar\Lambda^2<s_H<\bar\Lambda m_B$.

Although ${\rm d}\Gamma/{\rm d}s_H$ is integrable in perturbation theory,
powers of $\alpha_s\ln^2[(s_H-\bar\Lambda m_B)/m_B^2]$ occur in the invariant
mass spectrum.  This shows that perturbative and nonperturbative corrections
are both important for $s_H \lesssim \bar\Lambda m_B$.  (In the $m_b\to\infty$
limit perturbative corrections are important in a slightly larger region since
$\alpha_s\ln(s_H/m_B^2) \sim 1$ for $s_H \sim \bar\Lambda m_B$.)  While ${\rm
d}\Gamma/{\rm d}s_H$ cannot be reliably predicted for $s_H \lesssim \bar\Lambda
m_B$, the behavior of the spectrum for $s_H\lesssim\bar\Lambda m_B$ becomes
less important for observables that average over larger regions of the
spectrum, such as ${\rm d}\Gamma/{\rm d}s_H$ integrated over $s_H<\Delta^2$,
with $\Delta^2$ significantly greater than $\bar\Lambda m_B$.  

In Fig.~1 we plot the quantity $\hat\Gamma(\Delta^2,\bar\Lambda)$ 
defined by \cite{FLW}
\begin{equation}\label{Ghatdef}
\int_0^{\Delta^2} {\rm d}s_H\,
  {{\rm d}\Gamma(B\to X_u\,e\,\bar\nu)\over {\rm d}s_H} =
  {G_F^2\,m_B^5\over 192\pi^3}\, |V_{ub}|^2\,
  \bigg(1-{\bar\Lambda\over m_B}\bigg)^5\, \hat\Gamma(\Delta^2,\bar\Lambda) \,,
\end{equation}
as a function of $\Delta^2$ for $\bar\Lambda=0.2$, $0.4$ and $0.6\,$GeV in the
range $\bar\Lambda m_B<\Delta^2<4.5\,{\rm GeV}^2$, including terms up to order
$\alpha_s^2\beta_0$ (using $\alpha_s(m_b)=0.2$).  These curves approach
$\hat\Gamma(m_B^2,\bar\Lambda)\simeq0.73$ as $\Delta^2\to m_B^2$~\cite{LuSW}. 
$\hat\Gamma(\Delta^2,\bar\Lambda)/\hat\Gamma(m_B^2,\bar\Lambda)$ is the
fraction of events with hadronic invariant mass less than $\Delta^2$.  It is
mostly the ability to compute $\hat\Gamma(\Delta^2,\bar\Lambda)$ and our
knowledge of the value of $\bar\Lambda$ which determine the uncertainty from
theory in a value of $|V_{ub}|$ extracted from the invariant mass spectrum in
the region $s_H<\Delta^2$.

\begin{figure}[t]
\centerline{\epsfysize=8truecm \epsfbox{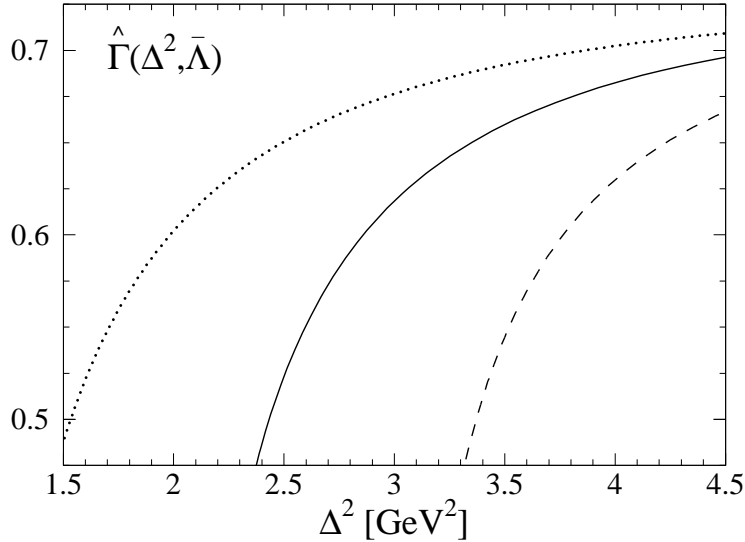}}
\caption[2]{The function $\hat\Gamma(\Delta^2,\bar\Lambda)$ defined in
Eq.~(\ref{Ghatdef}) as a function of $\Delta^2$ for $\bar\Lambda=0.2\,$GeV
(dotted curve), $0.4\,$GeV (solid curve), and $0.6\,$GeV (dashed curve).
Note that $m_D^2=3.5\,{\rm GeV}^2$.}
\label{fig:Gammahat}
\end{figure}

In the low mass region, $s_H\lesssim\bar\Lambda m_B$, nonperturbative
corrections from higher dimension operators in the OPE are very important. 
Just as in the case of the electron spectrum in the endpoint
region~\cite{shape}, the most singular terms can be identified and summed into
a shape function, $S(s_H)$.  Neglecting perturbative QCD corrections, we write
\begin{equation}\label{10}
{{\rm d}\Gamma\over {\rm d}s_H} = {G_F^2\,m_b^5\over192\pi^3}\,
  |V_{ub}|^2\, S(s_H)\,.
\end{equation}
It is convenient to introduce the scaled variable $y=s_H/\bar\Lambda m_b$.
Then \cite{FLW}
\begin{equation}\label{11}
\hat S(y) = \sum_{n=0}^\infty {(-1)^n A_n\over n!\,\bar\Lambda^n}\,
  {{\rm d}^n\over {\rm d}y^n} \left[2 y^{n+2} \left(3 - 2y\right)
  \theta(1-y) \right] ,
\end{equation}
where $\hat S(y)=\bar\Lambda m_b\,S(s_H)$ is dimensionless.  The matrix
elements $A_n$ are the same ones that determine the shape functions for the
semileptonic $B$ decay electron energy spectrum in the endpoint region, and 
also the photon energy endpoint region in weak radiative $B$ decay,
\begin{equation}\label{12}
\langle B(v)|\, \bar h_v^{(b)}\, iD_{\mu_1} \ldots iD_{\mu_n}\, h_v^{(b)}\,
  |B(v)\rangle/2m_B = A_n\, v_{\mu_1} \ldots v_{\mu_n} 
+ ({\rm terms~involving~} g_{\mu_i\mu_j}) \,.
\end{equation}
The $A_n$'s have dimension of $[\mbox{\it mass}]^n$, and hence the coefficients
$A_n/\bar\Lambda^n$ are dimensionless numbers of order one.  The first few
$A_n$'s are $A_0=1$, $A_1=0$, $A_2=-\lambda_1/3$, etc.

The shape function $\hat S(y)$ is an infinite sum of singular terms which gives
an invariant mass spectrum that leaks out beyond $y=1$ (i.e., $s_H =
\bar\Lambda m_b$).  For $y\sim1$, all terms in Eq.~(\ref{11}) are formally of
equal importance.  Since $\bar\Lambda m_b \sim 2\,{\rm GeV}^2$ is not too far
from $m_D^2$, it is necessary to estimate the influence of the nonperturbative
effects on the fraction of $B$ decays with invariant hadronic mass squared less
than $\Delta^2$.  It is difficult to obtain a model-independent estimate of the
leakage of events above an experimental cutoff $s_H=\Delta^2$, given that we
can estimate only the first few moments, $A_n$.  However, in the ACCMM model
\cite{ACCMM} with reasonable parameters, the shape function $\hat S(y)$ causes
a small (i.e., $\sim4\%$ with $\bar\Lambda=0.4\,$GeV, and perturbative QCD
corrections neglected) fraction of the events to have $s_H>m_D^2$ \cite{FLW}. 
Moreover, this leakage depends primarily on $\bar\Lambda$, and only to a lesser
extent on other ingredients of the model.
 
Thus the analysis of both perturbative and nonperturbative corrections implies
that the uncertainty in the determination of $|V_{ub}|$ from the hadronic
invariant mass spectrum in the region $s_H<m_D^2$ is largely controlled by the
uncertainty in $\bar\Lambda$, or equivalently, by that in the $b$ quark mass. 
(There is a so-called renormalon ambiguity in $\bar\Lambda$ and in the $b$
quark pole mass.  For the physically measurable quantity in
Eq.~(\ref{Ghatdef}), this is cancelled by a similar ambiguity in the
perturbative series in $\hat\Gamma$.)  To measure $|V_{ub}|$ with a theoretical
uncertainty below $\sim10\%$, $\bar\Lambda$ has to be determined
\cite{lambdabar} (with better than $\sim100\,$MeV uncertainty), and the cut on
$s_H$ has to be as close to $m_D^2$ as possible.  If the experimental
resolution forces one to consider a significantly smaller region then the
theoretical uncertainties will be larger.

\section{$V_{\lowercase{ub}}$ from exclusive $B$ decays}

Heavy quark symmetry \cite{HQS} is much less predictive for heavy to light
decays than it is for heavy to heavy transitions.  In the infinite mass limit
not all form factors are related to one another, and their normalization is not
fixed at any kinematic point.  There are still relations between semileptonic
$B$ and $D$ decays to the same charmless exclusive final state \cite{IsWi},
such as between $B\to\rho\,\ell\,\bar\nu$ and $D\to\rho\,\ell\,\bar\nu$, or
between $B\to\pi\,\ell\,\bar\nu$ and $D\to\pi\,\ell\,\bar\nu$.  The order
$1/m_{c,b}$ corrections to the infinite mass limit may be sizable, however, and
for final state pions there are additional complications since $m_\pi$ is
comparable to the mass difference between the vector and pseudoscalar mesons
\cite{Bpi}.  The question is whether we can construct an observable sensitive
to $V_{ub}$ that is (almost) free of $1/m_{c,b}$ corrections.

The basic idea \cite{lw,IsWi} is to use heavy quark symmetry to relate the
$SU(3)$ violation between $D\to K^*\,\bar\ell\,\nu$ and the Cabibbo suppressed
decay $D\to\rho\,\bar\ell\,\nu$ to those that occur in a comparison of $B\to
K^*\ell\,\bar\ell$ (or $B\to K^*\,\nu\,\bar\nu$) with $B \to
\rho\,\ell\,\bar\nu$.  Then experimental data on $B\to K^*\ell\,\bar\ell$ in
conjunction with data on $D\to\rho\,\bar\ell\,\nu$ and $D\to
K^*\,\bar\ell\,\nu$ can be used to determine $|V_{ub}|$.  This proposal is
complementary to other approaches for determining $|V_{ub}|$, since it relies
on the standard model correctly describing the rare flavor changing neutral
current process $B\to K^*\ell\,\bar\ell$. 

We denote by $g^{(H\to V)}$, $f^{(H\to V)}$, and $a_\pm^{(H\to V)}$ the form
factors relevant for semileptonic transitions between a pseudoscalar meson
containing a heavy quark $H$ ($H=B,D$), and a member of the lowest lying
multiplet of vector mesons $V$ ($V=\rho,K^*,\omega$), 
\begin{eqnarray}\label{ffdef}
\langle V(p',\epsilon) |\,\bar q\,\gamma_\mu\, Q\,| H(p)\rangle
&=& i\,g^{(H\to V)}\, \varepsilon_{\mu\nu\lambda\sigma}\, \epsilon^{*\nu}\,
  (p+p')^\lambda\, (p-p')^\sigma \,, \\*
\langle V(p',\epsilon) |\,\bar q\,\gamma_\mu\gamma_5\, Q\,| H(p)\rangle
&=& f^{(H\to V)}\,\epsilon^*_\mu 
  + a_+^{(H\to V)}\,(\epsilon^*\cdot p)\,(p+p')_\mu 
  + a_-^{(H\to V)}\,(\epsilon^*\cdot p)\,(p-p')_\mu \nonumber\,.
\end{eqnarray}
We view the form factors as functions of the dimensionless variable $y=v\cdot
v'$, where $p=m_H\,v$, $p'=m_V\,v'$, and $q^2 = (p-p')^2 = m_H^2 +m_V^2
-2m_H\,m_V\,y$.  (Although we are using the variable $v\cdot v'$, we are not
treating the quarks in $V$ as heavy.)  Assuming nearest pole dominance for the
$q^2$ dependences, the $D\to K^*\,\bar\ell\,\nu$ form factors are \cite{E791b}
\begin{eqnarray}\label{ffexp}
f^{(D\to K^*)}(y) &=& {(1.9\pm0.1)\,{\rm GeV}\over 1+0.63\,(y-1)}\,, 
  \nonumber\\*
a_+^{(D\to K^*)}(y) &=& -{(0.18\pm0.03)\,{\rm GeV}^{-1}\over 1+0.63\,(y-1)}\,, 
  \nonumber\\*
g^{(D\to K^*)}(y) &=& -{(0.49\pm0.04)\,{\rm GeV}^{-1}\over 1+0.96\,(y-1)}\,.
\end{eqnarray}
The shapes of these form factors are beginning to be probed experimentally
\cite{E791b}.  The form factor $a_-$ is not measured because its contribution
to the $D\to K^*\,\bar\ell\,\nu$ decay amplitude is suppressed by the lepton
mass.  These form factors are measured over the kinematic region
$1<y<(m_D^2+m_{K^*}^2)/(2m_D\,m_{K^*})\simeq1.3$.  Note that $f(y)$ changes by
less than 20\% over this range.  The full kinematic region for
$B\to\rho\,\ell\,\bar\nu$ is much larger, $1<y<3.5$.  In the following analysis
we will extrapolate the measured $D\to K^*$ form factors to $1<y<1.5$.  (The
validity of this extrapolation can be tested \cite{Boyd}.)

The differential decay rate for semileptonic $B$ decay (neglecting the lepton
mass, and not summing over the lepton type $\ell$) is
\begin{equation}\label{SLrate}
{{\rm d}\Gamma(B\to\rho\,\ell\,\bar\nu)\over{\rm d}y} 
  = {G_F^2\,|V_{ub}|^2\over48\,\pi^3}\, m_B\, m_\rho^2\, S^{(B\to\rho)}(y) \,.
\end{equation}
Here $S^{(H\to V)}(y)$ is the function
\begin{eqnarray}\label{shape}
S^{(H\to V)}(y) &=& \sqrt{y^2-1}\, \bigg[ \Big|f^{(H\to V)}(y)\Big|^2\,
  (2+y^2-6yr+3r^2) \nonumber\\*
&&\phantom{} + 4{\rm Re} \Big[a_+^{(H\to V)}(y)\, f^{(H\to V)}(y)\Big]
  m_H^2\, r\, (y-r) (y^2-1) \nonumber\\*
&&\phantom{} + 4\Big|a_+^{(H\to V)}(y)\Big|^2 m_H^4\, r^2 (y^2-1)^2 + 
  8\Big|g^{(H\to V)}(y)\Big|^2 m_H^4\, r^2 (1+r^2-2yr)(y^2-1)\, \bigg] 
  \nonumber\\*
&=& \sqrt{y^2-1}\, \Big|f^{(H\to V)}(y)\Big|^2\, (2+y^2-6yr+3r^2)\, 
  [1+\delta^{(H\to V)}(y)] \,,
\end{eqnarray}
with $r=m_V/m_H$.  The function $\delta^{(H\to V)}$ depends on the ratios of
form factors $a_+^{(H\to V)}/f^{(H\to V)}$ and $g^{(H\to V)}/f^{(H\to V)}$. 
$S^{(B\to\rho)}(y)$ can be estimated using combinations of $SU(3)$ flavor
symmetry and heavy quark symmetry.  $SU(3)$ symmetry implies that the $\bar
B^0\to\rho^+$ form factors are equal to the $B\to K^*$ form factors and the
$B^-\to\rho^0$ form factors are equal to $1/\sqrt2$ times the $B\to K^*$ form
factors.  Heavy quark symmetry implies the relations \cite{IsWi}
\begin{equation}\label{BDrel}
(f,\,a_+,\,g)^{(B\to K^*)} = \left({m_B\over m_D}\right)^{1/2} 
  \bigg[{\alpha_s(m_b)\over \alpha_s(m_c)}\bigg]^{-6/25}\, 
  (f,\,a_+,\,g)^{(D\to K^*)} \,,
\end{equation}
where we used $a_-^{(D\to K^*)}=-a_+^{(D\to K^*)}$, 
valid in the large $m_c$ limit.

Using Eq.~(\ref{BDrel}) and $SU(3)$ symmetry to get $\bar B^0\to \rho^+\,
\ell\,\bar\nu$ form factors (in the region $1<y<1.5$, corresponding to
$q^2>16\,{\rm GeV}^2$) from those for $D\to K^*\bar\ell\,\nu$ given in
Eq.~(\ref{ffexp}) yields $S^{(B\to\rho)}(y)$ plotted in Fig.~1 of
Ref.~\cite{lw}.  (The numerical values in Eq.~(\ref{ffexp}) differ slightly
from those used in Ref.~\cite{lw}.)  This prediction for $S^{(B\to\rho)}$ can
be used to determine $|V_{ub}|$ from the $B\to\rho\,\ell\,\bar\nu$ semileptonic
decay rate in the region $1<y<1.5$.  We find that about 20\% of $\bar
B^0\to\rho^+\ell\,\bar\nu$ decays are in the range $1<y<1.5$, and ${\cal B}
(B^0\to\rho^+ \ell\,\bar\nu)\Big|_{y<1.5} = 5.9\,|V_{ub}|^2$.  This method is
model independent, but cannot be expected to yield a very accurate value of
$|V_{ub}|$.  Typical $SU(3)$ violations are at the $10-20$\% level and similar
violations of heavy quark symmetry are expected.  In this region
$|\delta^{(B\to\rho)}(y)|$ defined in Eq.~(\ref{shape}) is less than 0.06,
indicating that $a_+^{(B\to\rho)}$ and $g^{(B\to\rho)}$ make only a small
contribution to the differential rate in this region.  Thus the main
uncertainties are $SU(3)$ and heavy quark symmetry violations in the $f^{(H\to
V)}$ form factor only; these are precisely the ones we can eliminate.

Ref.~\cite{lw} proposed a method for getting a value of $S^{(B\to\rho)}(y)$
with small theoretical uncertainty using a ``Grinstein-type" \cite{Gtdr} 
double ratio
\begin{equation}\label{Gtdr}
R(y) = \Big[ f^{(B\to\rho)}(y) / f^{(B\to K^*)}(y) \Big] \Big/
  \Big[ f^{(D\to\rho)}(y) / f^{(D\to K^*)}(y) \Big] \,,
\end{equation}
which is unity in the limit of $SU(3)$ symmetry or in the limit of heavy quark
symmetry.  Corrections to the prediction $R(y)=1$ are suppressed by
$m_s/m_{c,b}$ ($m_{u,d} \ll m_s$) instead of $m_s/\Lambda_{\rm QCD}$ or
$\Lambda_{\rm QCD}/m_{c,b}$.  The leading deviation of $R(1)$ from unity
arising from the Feynman diagrams in Fig.~2 has been estimated using chiral
perturbation theory.  These yield a calculable non-analytic $\sqrt{m_q}$
dependence on the light quark masses, and such corrections cannot arise from
other sources.  The result is $R(1)=1-0.035\,g\,g_2$ \cite{LSW}, where $g$ is
the $DD^*\pi$ coupling and $g_2$ is the $\rho\,\omega\,\pi$ coupling. 
Experimental data on $\tau\to\omega\,\pi\,\nu_\tau$ decay gives $g_2\simeq0.6$
\cite{dw}.  Estimates of $g$ vary between near unity and much smaller values
\cite{IWS}.  There may be significant corrections to $R(1)$ from higher orders
in chiral perturbation theory.  However, the smallness of our result lends
support to the expectation that $R(1)$ is very close to unity.  There is no
reason to expect any different conclusion over the kinematic range $1<y<1.5$.  

\begin{figure}[tb]  
\centerline{\epsfysize=2truecm \epsfbox{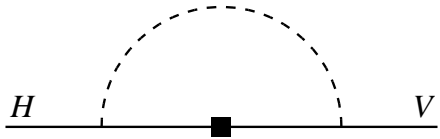}}
\caption[1]{Feynman diagram that gives the leading contribution to $R(1)-1$.
The dashed line is a $\pi$ or an $\eta$.
The black square indicates insertion of the weak current.}
\end{figure} 

Since $R(y)$ is very close to unity, the  relation 
\begin{equation}\label{magic}
S^{(B\to\rho)}(y) = S^{(B\to K^*)}(y)\,
  \bigg|{f^{(D\to\rho)}(y)\over f^{(D\to K^*)}(y)}\bigg|^2\,
  \bigg({m_B-m_\rho\over m_B-m_{K^*}}\bigg)^2\,,
\end{equation}
and measurements of $|f^{(D\to K^*)}|$, $|f^{(D\to\rho)}|$, and $S^{(B\to
K^*)}$ will determine $S^{(B\to\rho)}$ with small theoretical uncertainty.  The
last term on the right hand side makes Eq.~(\ref{magic}) equivalent to
Eq.~(\ref{Gtdr}) in the $y\to1$ limit.  The ratio of the
$(2+y^2-6yr+3r^2)\,[1+\delta^{(B\to V)}(y)]$ terms makes only a small and
almost $y$-independent contribution to $S^{(B\to\rho)}/S^{(B\to K^*)}$ in the
range $1<y<1.5$.  Therefore, corrections to Eq.~(\ref{magic}) are at most a few
percent larger than those to $R(y)=1$.

$|f^{(D\to K^*)}|$ has already been determined.  $|f^{(D\to\rho)}|$ may be
obtainable in the future, for example from experiments at $B$ factories, where
improvements in particle identification help reduce the background from the
Cabibbo allowed decay.  The measurement ${\cal B}(D\to\rho^0\,\bar\ell\,\nu) /
{\cal B}(D\to\bar K^{*0}\,\bar\ell\,\nu) = 0.047\pm0.013$ \cite{E791a} already
suggests that $|f^{(D\to\rho)}/f^{(D\to K^*)}|$ is close to unity.  Assuming
$SU(3)$ symmetry for the form factors, but keeping the explicit
$m_V$-dependence in $S^{(D\to V)}(y)$ and in the limits of the $y$ integration,
the measured form factors in Eq.~(\ref{ffexp}) imply ${\cal B}(D\to
\rho^0\,\bar\ell\,\nu) / {\cal B}(D\to \bar K^{*0}\,\bar\ell\,\nu) = 0.044$
\cite{LSW}.

$S^{(B\to K^*)}$ is obtainable from experimental data on $B\to K^*
\ell\,\bar\ell$ or $B\to K^*\, \nu\,\bar\nu$.  While the latter process is very
clean theoretically, it is very difficult experimentally.  A more realistic
goal is to use $B\to K^*\ell\,\bar\ell$, since CDF expects to observe
$400-1100$ events in the Tevatron Run~II (if the branching ratio is in the
standard model range) \cite{CDF2}.  The uncertainties associated with long
distance nonperturbative strong interaction physics in this extraction of
$S^{(B\to K^*)}(y)$, averaged over the region $1<y<1.5$, are probably less than
10\% \cite{LSW}.  Consequently, a determination of $|V_{ub}|$ from experimental
data on $D\to K^*\bar\ell\,\nu$, $D\to\rho\,\bar\ell\,\nu$, $B\to
K^*\ell\,\bar\ell$ and $B\to\rho\,\ell\,\bar\nu$ with an uncertainty from
theory of about 10\% is feasible.  If a precise value of $|V_{ub}|$ is
available before $B\to K^*\ell\,\bar\ell$ is measured, then we get an accurate
standard model prediction for the $B\to K^*\ell\,\bar\ell$ decay rate in the
region $1<y<1.5$.  Comparison with data may signal new physics or provide
stringent constraints on extensions of the standard model.

\section{Conclusions}

The present determinations of $|V_{ub}|$ rely on comparing experimental data
with model calculations, and therefore suffer from theoretical uncertainties of
order $30\%$.  (This is hard to quantify, and such a number is necessarily ad
hoc.)  To reduce these uncertainties one needs to consider somewhat different
observables for which the theoretical predictions are less model dependent than
those for the endpoint region of the inclusive electron spectrum and for the
total exclusive $B\to\pi\,\ell\,\bar\nu$ or $B\to\rho\,\ell\,\bar\nu$ decay
rates.  In this talk I reviewed two ideas which seem promising to me: i)
extracting $|V_{ub}|$ from the hadronic invariant mass spectrum in inclusive
semileptonic $B$ decays; and ii) using heavy quark and chiral symmetries for
form factors of exclusive semileptonic $B$ and $D$ decays to vector mesons. 
These may lead to model independent determinations of $|V_{ub}|$ with an
uncertainty from theory of about 10\%.  Lattice calculations \cite{lattice} and
dispersion relation constraints on form factors \cite{Boyd} will also be
important.

There is not really one ``gold-plated" observable for extracting $|V_{ub}|$. 
To reduce the strong interaction model dependence, several measurements will be
needed to guide us which approximations and expansions have smaller
uncertainties.  At the 10\% level consistency between different determinations
of $|V_{ub}|$ will be necessary to have confidence that the uncertainties are
indeed so small.  I am hopeful that this will be achieved within the next few
years.

\acknowledgements

I am grateful to Adam Falk, Iain Stewart, and Mark Wise for collaboration on
the topics discussed in this talk.  I thank the organizers for the invitation,
and for putting together a very interesting and enjoyable conference.  This
work was supported in part by the U.S.\ Dept.\ of Energy under grant no.\
DOE-FG03-97ER40506 and by NSF grant PHY-9457911.


\begin{references}

\bibitem{MaTr}
M. Trodden, these proceedings [hep-ph/9805252], and references therein.

\bibitem{incl}
F. Bartelt {\it et al.}, CLEO Collaboration, Phys. Rev. Lett. 71 (1993) 4111;\\
H. Albrecht {\it et al.}, Argus Collaboration, Phys. Lett. B255 (1991) 297.

\bibitem{excl}
J. Alexander {\it et al.}, CLEO Collaboration, 
Phys. Rev. Lett. 77 (1996) 5000.

\bibitem{CGG}
J. Chay {\it et al.}, Phys. Lett. B247 (1990) 399.

\bibitem{Bigi}
I.I. Bigi {\it et al.}, Phys. Lett. B293 (1992) 430,
I.I. Bigi {\it et al.}, Phys. Rev. Lett. 71 (1993) 496.

\bibitem{MaWi}
A.V. Manohar and M.B. Wise, Phys. Rev. D49 (1994) 1310; 
B. Blok {\it et al.}, Phys. Rev. D49 (1994) 3356
  [(E) {\it ibid.} D50 (1994) 3572]; 
T. Mannel, Nucl. Phys. B413 (1994) 396.

\bibitem{FLW}
A.F. Falk {\it et al.}, Phys. Lett. B406 (1997) 225.

\bibitem{BDU}
R.D. Dikeman and N. Uraltsev, Nucl. Phys. B509 (1998) 378; \\
I. Bigi {\it et al.}, TPI-MINN-97-21-T [hep-ph/9706520].

\bibitem{old}
V. Barger {\it et al.}, Phys. Lett. B251 (1990) 629; 
J. Dai, Phys. Lett. B333 (1994) 212.

\bibitem{ISGW}
N. Isgur {\it et al.}, Phys. Rev. D39 (1989) 799; \\
N. Isgur and D. Scora, Phys. Rev. D52 (1995) 2783.

\bibitem{LuSW}
M. Luke {\it et al.}, Phys. Lett. B343 (1995) 329.

\bibitem{shape}
M. Neubert, Phys. Rev. D49 (1994) 3392; D49 (1994) 4623; \\
I.I. Bigi {\it et al.}, Int. J. Mod. Phys. A9 (1994) 2467.

\bibitem{ACCMM}
G. Altarelli {\it et al.}, Nucl. Phys. B208 (1982) 365; \\
A. Ali and I. Pietarinen, Nucl. Phys. B154 (1979) 519.

\bibitem{lambdabar}
M. Gremm {\it et al.}, Phys. Rev. Lett. 77 (1996) 20;
A.F. Falk {\it et al.}, Phys. Rev. D53 (1996) 6316;
A. Kapustin and Z. Ligeti, Phys. Lett. B355 (1995) 318; 
Z. Ligeti and Y. Nir, Phys. Rev. D49 (1994) 4331;
M. Luke and M.J. Savage, Phys. Lett. B321 (1994) 88.

\bibitem{HQS}
N. Isgur and M.B. Wise, Phys. Lett. B232 (1989) 113; 
Phys. Lett. B237 (1990) 527.

\bibitem{IsWi}
N. Isgur and M.B. Wise, Phys. Rev. D42 (1990) 2388.

\bibitem{Bpi}
N. Isgur and M.B. Wise, Phys. Rev. D41 (1990) 151; 
M.B. Wise, Phys. Rev. D45 (1992) 2188; 
G. Burdman and J.F. Donoghue, Phys. Lett. B280 (1992) 287; 
L. Wolfenstein, Phys. Lett. B291 (1992) 177; 
G. Burdman {\it et al.}, Phys. Rev. D49 (1994) 2331.

\bibitem{lw}
Z. Ligeti and M.B. Wise, Phys. Rev. D53 (1996) 4937. 

\bibitem{E791b}
E.M. Aitala {\it et al.}, E791 Collaboration, Phys. Rev. Lett. 80 (1998) 1393.

\bibitem{Boyd}
See, e.g., C.G. Boyd {\it et al.}, Phys. Rev. Lett. 74 (1995) 4603; 
C.G. Boyd and I.Z. Rothstein, Phys. Lett. B420 (1998) 350; 
C.G. Boyd and M.J. Savage, Phys. Rev. D56 (1997) 303;
L.~Lellouch, Nucl. Phys. Proc. Suppl. 54A (1997) 266;
G. Burdman and J. Kambor, Phys. Rev. D55 (1997) 2817. 

\bibitem{Gtdr}
B. Grinstein, Phys. Rev. Lett. 71 (1993) 3067.

\bibitem{LSW}
Z. Ligeti {\it et al.}, Phys. Lett. B420 (1998) 359.

\bibitem{dw}
H. Davoudiasl and M.B. Wise, Phys. Rev. D53 (1996) 2523.

\bibitem{IWS}
I.W. Stewart, CALT-68-2160 [hep-ph/9803227], and references therein.

\bibitem{E791a}
E.M. Aitala {\it et al.}, E791 Collaboration, Phys. Lett. B397 (1997) 325.

\bibitem{CDF2}
CDF Collaboration, CDF~II Detector Technical Design Report,
Fermilab-Pub-96/390-E.

\bibitem{lattice}
For a recent review see, 
J.M. Flynn and C.T. Sachrajda, SHEP-97-20 [hep-lat/9710057]; 
it may be possible to measure the shape function directly on the lattice, see,
U. Aglietti {\it et~al.}, ROME1-1204/98 [hep-ph/9804416].

\end{references}
\end{document}